\documentclass[12pt,a4paper]{article}
\usepackage[]{epsfig}	      
\begin{document}
 
 \thispagestyle{empty}
 
\date{October 7, 2001}
 \title{The double torus as a 2D cosmos:\\
 Groups, geometry, and closed geodesics.}
 \author{Peter Kramer$^a$ and Miguel Lorente$^b$,\\
\small $^a$ Institut f\"ur Theoretische Physik der 
 Universit\"at D 72076 T\"ubingen, Germany,\\
 \small $^b$ Departamento de Fisica, Universidad de Oviedo, E 33007 Oviedo, Spain.
 \\{\em \small Dedicated to Marcos Moshinsky on the occasion of his 80th birthday}}
\maketitle
 
\section*{Abstract.}
The double torus provides a relativistic model for a closed 
2D cosmos with topology of genus 2 and 
constant negative
curvature. Its unfolding into an octagon extends to an octagonal 
tesselation of its universal covering, the hyperbolic space
$H^2$. The tesselation is analysed with tools from hyperbolic 
crystallography. Actions on $H^2$ of groups/subgroups  are identified for $SU(1,1)$, 
for a hyperbolic Coxeter group acting also on $SU(1,1)$, and for the 
homotopy group $\Phi_2$ whose extension is normal in the Coxeter group.
Closed geodesics arise from links  on $H^2$ between octagon centers.
The direction and length of the shortest closed geodesics is computed.

\section{Introduction.}
In a publication \cite{LA} entitled Cosmic Topology, M. Lachieze-Rey and 
J.-P. Luminet review topological alternatives for
pseudoriemannian manifolds of constant negative curvature 
as cosmological models and discuss their
implications for observations. The general classification of spaces of constant 
curvature was given by Wolf \cite{WO}. Some of these  manifolds 
admit closed (null-) geodesics.
Test particles (photons) travelling along such geodesics produce  images of their 
own source and therefore observable effects under conditions described
in \cite{LA} pp. 189-202. 
All these simple geometric models of a cosmos neclect the influence of 
varying mass  distributions. These distort locally the curvature and 
the   geodesics assumed in the models for test particles, 
compare \cite{LA} p. 173.
Of particular interest for observing the topology, 
compare \cite{LA} for details, are the shortest closed geodesics in 
at least three ways: 
(i) They produce the brightest images of a source and  are 
considered less sensitive to the local distortions. (ii) They determine
the first peaks in the autocorrelation function of the matter distribution.
(iii) By their length they determine a  global radius
of the model. If no correspondence between sources and images can be observed,
this puts a lower limit on this global radius. 
\vspace{0.2cm} 

For comparison we sketch  an analysis for the   
2D torus manifold with
Riemannian metric. Its fundamental polygon can be taken as a unit square.
Its universal covering manifold is a plane which admits a tesselation 
by copies of one  fundamental initial square. 
The Euclidean metric on the plane $E^2$ induces on the 
torus a Riemannian metric with  zero curvature. The homotopy group 
of the torus is $Z \times Z$. This homotopy group is isomorphic
to a 2D translation group acting on the plane whose elements $(m, n)$ 
yield the square lattice of center
positions of all squares, seen from the initial square.
The straight lines on the plane, when  pulled back to the reference square, 
determine its geodesics. 
Consider now a
straight line passing through the center of the initial square.
If and only if it hits for the first time the center $(m,n)$ of another square,
then $m,n$ are relatively prime, and the 
geodesic when pulled back to the initial square will close 
on the torus. The slope  of this closed geodesic is 
$n/m$, its length is $s=\sqrt{m^2+n^2}$. The problem 
of closed geodesics on the torus is thus solved by simple crystallographic 
computations on the plane. 
These relations are illustrated in Fig. 1. The shortest 
closed geodesics of length  $s=1$ clearly run between the centers 
of the initial square and its $4$ edge neighbours. By giving the 
squares a general fixed edge length one could introduce a global radius
of this torus model. 
 
\begin{center}
\begin{picture}(0,0)%
\includegraphics{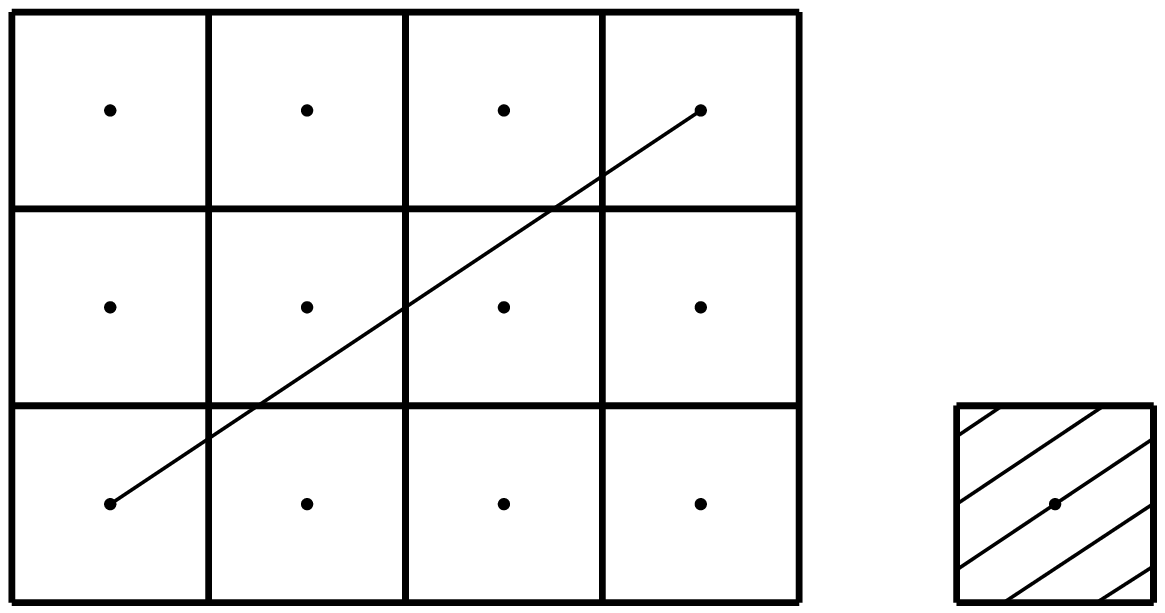}%
\end{picture}%
\setlength{\unitlength}{8287sp}%
\begingroup\makeatletter\ifx\SetFigFont\undefined%
\gdef\SetFigFont#1#2#3#4#5{%
  \reset@font\fontsize{#1}{#2pt}%
  \fontfamily{#3}\fontseries{#4}\fontshape{#5}%
  \selectfont}%
\fi\endgroup%
\begin{picture}(2654,1394)(429,817)
\end{picture}

\end{center}

Fig. 1. The universal covering manifold of the torus is a plane 
$E^2$ tesselated by unit squares. 
The homotopy group $Z \times Z$ of the torus
acts on $E^2$ by translations $(m,n)$. The Euclidean metric on $E^2$ induces
on the torus a Riemannian metric. Geodesic sections 
between centers of squares on $E^2$ become closed geodesics on the torus.
On the left, a geodesic 
line on $E^2$ of slope $2/3$ and of length
$s=\sqrt{3^2+2^2}$ connects the centers $(0,0)$ and $(3,2)$ on $E^2$. 
On the right, this geodesic is pulled back to the initial square.
\vspace{0.5cm}

In what follows we 
examine the  continuous and discrete groups
and the geometry for the 2D double torus as a 
paradigm for a cosmos with closed geodesics. 
The analysis will  in concept follow
similar steps as sketched above for the torus. 
In contrast and in detail it will
involve the action of crystallographic groups on hyperbolic space with
pseudoriemannian metric, a field in which the present authors
have been  interested  \cite{KR2,KR3}. 
We believe that similar 
techniques apply to some other topologies of \cite{LA}.

We start from the information on the double torus given in \cite{LA}. 
The fundamental domain of the double torus is a hyperbolic octagon. 
It was described by Hilbert and 
Cohn-Vossen  in 1932 \cite{HI}. 
The universal covering manifold of the double torus must admit 
a tesselation by these octagons. This topological condition 
enforces the hyperbolic space $H^2$ as 
the universal covering of the double torus. The homotopy group
of the double torus as discussed by Seifert and Threlfall \cite{SE}
pp. 6, 174
and by Coxeter \cite{CO} p. 59 can be converted  into a group $\Phi_2$ acting on
the universal covering manifold 
$H^2$. The generators and the single relation for this group 
were given by Coxeter \cite{CO} and Magnus \cite{MA}. 
On the universal covering manifold we shall
see that the search for pre-images of closed geodesics becomes a problem of
hyperbolic crystallography.
\vspace{0.2cm}

In section 2 we describe the group $SU(1,1)$, its action and relevant
cosets.
In section 3 we review the universal covering of the double torus,
the hyperbolic space  $H^2 \sim SU(1,1)/U(1)$. 
Equipped with a pseudoriemannian structure it 
has constant negative curvature and is invariant under the action of
$SU(1,1)$.
In section 4 we give the general group description and group action for 
$H^2$. 
The octagonal fundamental domain of the double torus is identified as a 
double coset of $SU(1,1)$ with respect to the left action of the 
homotopy group $\Phi_2$ and the 
right action of $U(1)$. In section 5 we analyze  the homotopy group 
as a subgroup
of a hyperbolic Coxeter group. An extension of the homotopy group
is shown to be a normal subgroup of this hyperbolic Coxeter group which by
conjugation yields the octagonal symmetry. 
In section 6 we turn to the geometric action of the homotopy group.
A geometric origin of the
relation between the generators is given.
We characterize  on $SU(1,1)/U(1)$ the condition for closed
geodesics. We determine the next neighbour octagons
in the tesselation and from them find the shortest closed geodesics on the 
double torus.

\section{Groups and actions on Minkowski space $M(1,2)$.}

Here we describe the group $SU(1,1)$, its universal covering 
relation to $SO_{\uparrow}(1,2)$, 
and their  action on Minkowski space
$M(1,2)$.

\subsection{The group $SU(1,1)$.}

We first discuss group elements $g$ of $SU(1,1)$ and their parameters.
\begin{eqnarray}
\label{g01}
g \in SU(1,1)&:& gMg^{\dagger}M=e, 
M= 
\left[
\begin{array}{ll}
1&0\\
0& -1
\end{array}
\right],
M^2=e,
\\ \nonumber
g &=& 
\left[
\begin{array}{ll}
\lambda&\mu\\
\overline{\mu}& \overline{\lambda}
\end{array}
\right], |\lambda|^2-|\mu|^2=1,
\\ \nonumber
\lambda&=& \xi_0+i\xi_1,\; \mu= \xi_3-i\xi_2.
\end{eqnarray}
Here $\xi_j$ are real parameters.

Exponential parameters arise as follows: Define in terms of the Pauli matrices
the $2\times 2$ matrix representation of the Lie algebra $su(1,1)$, 
\begin{eqnarray}
\label{g02}
\tau_0&:=& i\sigma_3, \tau_1:=\sigma_2, \tau_2:=\sigma_1,
\\ \nonumber
-\tau_0^2&=& \tau_1^2=\;\tau_2^2\;=1,
\\ \nonumber
g&=& \exp(\tilde{\alpha} h),\;
h+Mh^{\dagger}M=0,\; \tilde{\alpha}\; {\rm real},
\\ \nonumber
h&=& \sum_0^2 \omega_j \tau_j
\\ \nonumber
&=& 
\left[
\begin{array}{ll}
i\omega_0&\omega_2-i\omega_1\\
\omega_2+i\omega_1& -i\omega_0
\end{array}
\right],
\\ \nonumber
(\tilde{\alpha}h)^2&=&-(\tilde{\alpha})^2(\omega_0^2-\omega_1^2-\omega_2^2)e. 
\end{eqnarray}
With the exponential parameters, elements of $SU(1,1)$ are 
described by an  angle $\tilde{\alpha}$ and by the  vector 
$\omega=(\omega_0,\omega_1,\omega_2)$ in $M(1,2)$, compare eq. \ref{g07}. 
Depending on the value of 
$|tr(g)|/2=|(\lambda+\overline{\lambda})|/2:
\langle \; <1, >1, =1 \rangle$
we have three cases, the elliptic case
\begin{eqnarray}
\label{g03} 
1&=&(\omega_0^2-\omega_1^2-\omega_2^2),
\\ \nonumber
g&=& \cos(\tilde{\alpha})e 
+ \sin(\tilde{\alpha})
(\omega_0\tau_0+\omega_1\tau_1+\omega_2\tau_2),
\\ \nonumber
\xi_0&=&\cos(\tilde{\alpha}),\; \xi_j= \sin(\tilde{\alpha}) \omega_{j+1},
\end{eqnarray}
the hyperbolic case
\begin{eqnarray}
\label{g04}
-1&=&(\omega_0^2-\omega_1^2-\omega_2^2),
\\ \nonumber
g&=& \cosh(\tilde{\alpha}) 
+ \sinh(\tilde{\alpha})(\omega_0\tau_0+\omega_1\tau_1+\omega_2\tau_2),
\\ \nonumber 
\xi_0&=&\cosh(\tilde{\alpha}),\; \xi_j= \sinh(\tilde{\alpha}) \omega_{j+1},
\end{eqnarray}
and the Jordan case
\begin{eqnarray}
\label{g05} 
0&=&(\omega_0^2-\omega_1^2-\omega_2^2),
\\ \nonumber
g&=& e+ (\omega_0\tau_0+\omega_1\tau_1+\omega_2\tau_2),
\\ \nonumber
\xi_0&=& 1,\; \xi_j= \omega_{j+1},
\end{eqnarray}
where the real parameters $\xi_j$ of eq. \ref{g01} are 
identified in each case.

An element of elliptic type may be written, compare eqs. \ref{g09}
and \ref{g011}, as
\begin{eqnarray}
\label{g06}
g& =& r_3(\alpha)b_2(\theta)r_3(\gamma)
r_3(\tilde{\alpha})r_3(-\gamma)b_2(-\theta)r_3(-\delta)
\\ \nonumber
&=& \left[
\begin{array}{ll}
\cos(\tilde{\alpha})+i\cosh(2\theta)\sin(\tilde{\alpha})&
-i\sinh(2\theta)\sin(\tilde{\alpha})\exp(2i\alpha)\\
i\sinh(2\theta)\sin(\tilde{\alpha})\exp(-2i\alpha)&
\cos(\tilde{\alpha})-i\cosh(2\theta)\sin(\tilde{\alpha})
\end{array}
\right],
\\ \nonumber
&=&
 \left[
\begin{array}{ll}
\cos(\tilde{\alpha})+i\sin(\tilde{\alpha})\omega_0&
\sin(\tilde{\alpha})(\omega_2-i\omega_1)\\
\sin(\tilde{\alpha})(\omega_2+i\omega_1)&
\cos(\tilde{\alpha})-i\sin(\tilde{\alpha})\omega_1
\end{array}
\right],
\\ \nonumber
\omega&=&(\cosh(2\theta),\sinh(2\theta)\cos(2\alpha),
\sinh(2\theta)\sin(2\alpha)).
\end{eqnarray}

The adjoint action of $SU(1,1)$ on its Lie algebra can be written by
use of eq. \ref{g01} as
\begin{eqnarray}
\label{g061}
(g, h) &\rightarrow & ghg^{-1},
\\ \nonumber
(g, -ihM) &\rightarrow & g(-ih)g^{-1}M=g(-ihM)g^{\dagger},
\\ \nonumber
(-ihM)^{\dagger}&=& -ihM.
\end{eqnarray}
By the adjoint action, the vector components 
$\omega(g)=(\omega_0,\omega_1,\omega_2)$
associated with $g$  
are linearly transformed according to the homomorphism 
$SU(1,1) \rightarrow SO_{\uparrow}(1,2)$, see subsection 2.2.
Any vector $\omega(g)$ is fixed under the action of $g$. 
The vectors $\omega$ for elliptic, hyperbolic and Jordan case 
correspond to and transform as time-like, space-like and light-like vectors
on $M(1,2)$.

\subsection{Group action and coset spaces.}

We adopt a vector and matrix notation in Minkowski space $M(1,2)$:
\begin{eqnarray}
\label{g07}
x &=& (x_0, x_1, x_2),
\\ \nonumber 
\langle x,y\rangle := x_0y_0-x_1y_1-x_2y_2,
\\ \nonumber
\tilde{x} &:=& \left[
\begin{array}{ll}
x_0 & x_1+ix_2\\
x_1-ix_2& x_0
\end{array}
\right].
\end{eqnarray}
The group action on $M(1,2)$ for $g \in SU(1,1)$ we define by 
\begin{eqnarray}
\label{g08}
(g, \tilde{x}) &\rightarrow & g\tilde{x} g^{\dagger},
\\ \nonumber
(g, x_i)  &\rightarrow & 
(gx)_i = \sum_j L_{ij}(g)x_j.
\end{eqnarray}
This action preserves hermiticity, the determinant, 
and hence the metric on $M(1,2)$.
$L(g)$ with $L(-g)=L(g)$ is the two-to-one homomorphism  of
$SU(1,1)$ to $SO_{\uparrow}(1,2)$, the orthochronous Lorentz group, compare \cite{KR93}. 
We have chosen the action of $SU(1,1)$ on $M(1,2)$ in eq. \ref{g08}
to agree with the 
adjoint action  eq. \ref{g061} on the Lie algebra with the exponential
parameters eq. \ref{g02}.
We define the elements
\begin{eqnarray}
\label{g09}
r_3(\alpha) 
&=& \left[
\begin{array}{ll}
\exp(i\alpha) & 0\\
0 & \exp(-i\alpha)
\end{array}
\right],\: \omega=(1,0,0),
\\ \nonumber 
b_2(\theta) 
&=& \left[
\begin{array}{ll}
\cosh(\theta) & \sinh(\theta)\\
\sinh(\theta)& \cosh(\theta)
\end{array}
\right], \: \omega=(0,0,1),
\end{eqnarray}
of  $SU(1,1)$ whose homomorphic images
\begin{eqnarray}
\label{g010}
 L_3(2\alpha)=L(r_3(\alpha))
&=& \left[
\begin{array}{lll}
1 & 0&0\\
0 & \cos(2\alpha)& -\sin(2\alpha)\\
0 & \sin(2\alpha)& \cos(2\alpha)
\end{array}
\right],
\\ \nonumber 
L_2(2\theta)= L(b_2(\theta)) 
&=& \left[
\begin{array}{lll}
\cosh(2\theta) & \sinh(2\theta)&0\\
\sinh(2\theta)& \cosh(2\theta)&0\\
0&0&1
\end{array}
\right]
\end{eqnarray}
are a rotation from $U(1)$ or a boost respectively. The 
factor $2$ in the angular parameters reflects our emphasis
on the group $SU(1,1)$.
A general group element of $SU(1,1)$ admits the Euler-type
parametrization
\begin{equation}
\label{g011}
g = g(\Omega)= g(\alpha, \theta, \gamma):=r_3(\alpha)b_2(\theta)r_3(\gamma).
\end{equation}
We shall need the multiplication rule for two elements $g_1,g_2$
written in these parameters. A special case  of this 
multiplication is 
\begin{eqnarray}
\label{g012}
 g_1(0, \theta_1, \gamma_1) g_2(0, \theta_2, 0)
 &=&  g(\alpha, \theta, \gamma),
 \\ \nonumber
\cosh(2\theta)&=& \cosh(2\theta_1)\cosh(2\theta_2)
+\sinh(2\theta_1)\sinh(2\theta_2)\cos(2\gamma_1),
\\ \nonumber 
\sinh(2\theta)\sin(2\alpha)&=& \sinh(2\theta_2)\sin(2\gamma_1),
\\ \nonumber
\sinh(2\theta)\sin(2\gamma)&=& \sinh(2\theta_1)\sin(2\gamma_1).
\end{eqnarray}
For general products 
$g_1(\alpha_1,\theta_1,\gamma_1)g_2(\alpha_2,\theta_2,\gamma_2)$
we conclude from eq. \ref{g012} that
\begin{equation}
\label{g013}
\cosh(2\theta(g_1g_2))= \cosh(2\theta_1)\cosh(2\theta_2)
+\sinh(2\theta_1)\sinh(2\theta_2)\cos(2(\gamma_1+\alpha_2)).
\end{equation}
The general multiplication law is obtained from eq. \ref{g013} by  
right- and left- multiplications with rotation matrices of
type $r_3(\alpha)$.

The hyperbolic space 
$H^2$, compare section 3, is the coset space $SU(1,1)/U(1)$. 
It is a homogeneous 
space under the action of $SU(1,1)$. 
The points of 
$SU(1,1)/U(1)$ we choose as the elements  $c \in SU(1,1)$ whose Euler angle
parameters  from eq. \ref{g011} obey $\gamma(c) =0$. 
Any element $g \in SU(1,1)$ can now be written as
\begin{equation}
\label{g0132}
g= c\,h, \; h \in U(1).
\end{equation}
The vectors 
$x \in M(1,2),\; \langle x,x \rangle =1$ of $H^2$, and the coset elements 
$c$ are in one-to-one correspondence, $x  \leftrightarrow c$.

\subsection{Involutive automorphisms of $SU(1,1)$.}

There are two involutive automorphisms 
$\psi_1, \psi_2,\; \psi_1^2=e, \psi_2^2=e$ of $SU(1,1)$:
\begin{eqnarray}
\label{g015}
(\psi_1, g)&\rightarrow& \overline{g},
\\ \nonumber
(\psi_2, g)&\rightarrow& (g^{\dagger})^{-1}.
\end{eqnarray}
The action of these automorphisms is given explicitly by
\begin{eqnarray}
\label{g016}
\psi_1(
\left[
\begin{array}{ll}
\lambda&\mu\\
\overline{\mu}&\overline{\lambda}
\end{array}
\right])
&=&
\left[
\begin{array}{ll}
\overline{\lambda}&\overline{\mu}\\
\mu&\lambda\\
\end{array}
\right],
\\ \nonumber
\psi_2(
\left[
\begin{array}{ll}
\lambda&\mu\\
\overline{\mu}&\overline{\lambda}
\end{array}
\right])
&=&
\left[
\begin{array}{ll}
\lambda&-\mu\\
-\overline{\mu}&\overline{\lambda}
\end{array}
\right].
\end{eqnarray}
The stability subgroups $H_{\psi_i}$ under these automorphisms are easily 
seen to be $H_{\psi_1}= SO(1,1) =\langle b_2(\theta) \rangle$, 
$H_{\psi_2}= U(1)= \langle r_3(\alpha) \rangle$. 
It follows that the automorphisms must map the coset spaces 
with respect to these subgroups into themselves.

We shall combine these automorphisms together with the left action 
$(g_1,g) \rightarrow g_1g$ into a larger group. This larger group
admits the involutions 
$\psi_1(g_1):=g_1 \circ \psi_1 \circ g_1^{-1}, 
\psi_2(g_2):=g_2 \circ \psi_2 \circ g_2^{-1}$
which act on $SU(1,1)$ as
\begin{eqnarray}
\label{g017} 
(\psi_1(g_1),g)&\rightarrow& (g_1 \circ \psi_1 \circ g_1^{-1})g=
g_1(\overline{g}_1)^{-1}\overline{g},
\\ \nonumber
(\psi_2(g_2),g) &\rightarrow& (g_2 \circ \psi_2 \circ g_2^{-1})g=
g_2 g_2^{\dagger}(g^{\dagger})^{-1}.
\end{eqnarray}
Here and in what follows we use the symbol $\circ$ to emphasize
operator multiplication.
 From the stability groups for $\psi_1,\psi_2$
it follows that all $\psi_1(g_1),\psi_2(g_2)$ are 
in one-to-one correspondence to the points of
the cosets $g_1 \in SU(1,1)/SO(1,1), g_2 \in SU(1,1)/U(1)$.  

\subsection{Weyl reflections and involutive automorphisms.}

A Weyl reflection acting on  $M(1,2)$ with a Weyl vector $k$ 
is defined as  the involutive map
\begin{equation}
\label{g014}
(W_k,x) \rightarrow y= x-2(\langle x,k\rangle /\langle k,k\rangle) \;k. 
\end{equation}
It preserves the scalar product eq. \ref{g07}, and leaves any
point $x$ of the plane $\langle k, x \rangle=0$ fixed.
Weyl reflections act on the coset spaces in $M(1,2)$.
We can distinguish two types of Weyl reflections for 
$k$ time-like or space-like. All Weyl operators $W_k$ in $M(1,2)$ 
with $k$ space-like may be expressed with 
\begin{eqnarray}
\label{gA1}
k&=& L(r_3(\alpha))L(b
_2(\theta))(0,1,0)
\\ \nonumber 
&=&  L(r_3(\alpha))L(b
_2(\theta))L(r_3(\pi/4))(0,0,-1)
\\ \nonumber 
&=& (\sinh(2\theta), \cosh(2\theta)\cos(2\alpha), \cosh(2\theta)\sin(2\alpha)).
\end{eqnarray}
Here we let $L(g)$ represent actions $g \in SU(1,1)$ on $M(1,2)$. As representative 
it proves convenient to pass to $k^0=(0,0,-1)$ with  
\begin{equation}
\label{gA2}
(W_{(0,0,-1)}, (x_0,x_1,x_2)) \rightarrow (x_0,x_1,-x_2).
\end{equation}

For Weyl reflections we have the general conjugation property
\begin{equation}
\label{gA21}
L(g) \circ W_{k} \circ L(g^{-1})= W_{L(g)k}
\end{equation}
which applies in particular to $k^0=(0,0,-1)$.

We now wish to define operators acting on $SU(1,1)$ by left 
multiplication and by involutive automorphisms which correspond to Weyl
reflections.
In the disc model of section 3.2,  the Weyl reflection of eq. \ref{gA2} corresponds to complex conjugation, and we 
therefore shall associate with $k^0=(0,0,-1)$ the automorphism 
$\psi_1$ of eq. \ref{g017}.
For operator products we have the identity 
\begin{equation}
\label{gA22}
\psi_1 \circ g \circ \psi_1 = \overline{g}.
\end{equation}
It allows to convert operator products containing  an even number 
of automorphisms into pure left multiplications.
Guided by eqs. \ref{gA1}, \ref{gA21}
we define an involutive automorphism corresponding
to a general Weyl vector $k$ by forming in correspondence to
eq. \ref{gA21} the operator product
\begin{eqnarray}
\label{gA3}
s_k:& =& g \circ \psi_1 \circ g^{-1}
 =g \circ \psi_1 \circ g^{-1}\circ \psi_1 \circ \psi_1
= g (\overline{g})^{-1} \circ \psi_1,
\\ \nonumber
g&=& r_3(\alpha)b_2(\theta)r_3(\pi/4),
\\ \nonumber 
g(\overline{g})^{-1}
&=& r_3(\alpha)b_2(\theta)r_3(\pi/2)b_2(-\theta)r_3(\alpha)
\\ \nonumber
&=&
\left[
\begin{array}{ll}
i\cosh(2\theta)\exp(2i\alpha)& -i\sinh(2\theta)\\
i\sinh(2\theta)& -i\cosh(2\theta)\exp(-2i\alpha)
\end{array}
\right].
\end{eqnarray}
The matrix $g$ is taken from eq. \ref{gA1} but expressed 
on the level of $SU(1,1)$.

\section{The two-dimensional hyperbolic manifold $H^2$.}

In this section we introduce the hyperbolic manifold $H^2$ as 
a hyperboloid in $M(1,2)$, the hyperbolic disc model, and 
review the pseudoriemannian structure on $H^2$.  

\subsection{$H^2$ as a hyperboloid.}

In $M(1,2)$ we consider the hyperboloid
$H^2: \langle x,x \rangle = x_0^2-x_1^2-x_2^2=1,\; x_0>0$,
compare Ratcliffe \cite{RA} pp. 56-104.
Its representative point  $x^0=(1,0,0)$ has the stability group
$U(1)\sim SO(2,R)$ generated by rotations. Therefore  
the  hyperbolic manifold $H^2$ is the coset space $SU(1,1)/U(1)$.
A parametrization of $H^2$ is obtained from eqs. \ref{g07},\ref{g08}:
\begin{eqnarray}
\label{g20}
x &=& 
L(r_3(\alpha)b_2(\theta)) x^0,
\\ \nonumber 
&=&(\cosh(2\theta), \sinh(2\theta)\cos(2\alpha), \sinh(2\theta) \sin(2\alpha)),
\\ \nonumber
\tilde{x}&\rightarrow& \tilde{x}(\theta,\alpha) 
\\ \nonumber
&=&
r_3(\alpha)b_2(\theta)
\left[
\begin{array}{ll}
1 & 0\\
0 & 1
\end{array}
\right]
b_2^{\dagger}(\theta)r_3^{\dagger}(\alpha)
\\ \nonumber
&=& 
\left[
\begin{array}{ll}
\cosh(2\theta) & \sinh(2\theta)\exp(2i\alpha)\\
\sinh(2\theta)\exp(-2i\alpha) & \cosh(2\theta) 
\end{array}
\right].
\end{eqnarray}
The action of $SU(1,1)$ on $H^2$ is given in eqs. \ref{g07},\ref{g08}.
We introduce in $M(1,2)$ intersections of the coset space
$SU(1,1)/U(1)$ with planes through $(0,0,0)$ characterized by their 
normal vector $k$, $\langle k, x \rangle =0$. These intersections
exist only if $k$ is space-like. They 
always  define geodesics, compare \cite{RA} pp. 68-70.
For $k=(0,1,0)$ we can write the intersection points
as $(x_0,x_1,x_2)=(\cosh(2\rho),0, \sinh(2\rho))$. Then $2\rho$ is 
a geodesic length parameter, compare section 3.2. 
The application of $r_3(\alpha)b_2(\theta)$ to $k$ and $x$ 
from eqs. \ref{g09},\ref{g010} yields
\begin{eqnarray}
\label{g21}
k &\rightarrow& (\sinh(2\theta), \cosh(2\theta)\cos(2\alpha),
\cosh(2\theta)\sin(2\alpha)),
\\ \nonumber 
x(\rho)
&=&
(\cosh(2\theta)\cosh(2\rho),
\\ \nonumber 
&&\cos(2\alpha) \sinh(2\theta)\cosh(2\rho)-\sin(2\alpha)\sinh(2\rho),
\\ \nonumber
&&\sin(2\alpha) \sinh(2\theta)\cosh(2\rho)+\cos(2\alpha)\sinh(2\rho)).
\end{eqnarray}
All the vectors $k$ normal to planes which intersect $H^2$ are space-like 
and may be written in terms of $(\alpha,\beta)$. Geodesics are 
mapped into geodesics under $SU(1,1)$. There  follows 
 
{\bf Prop 1}: The expression eq. \ref{g21} yields the most general 
geodesics on $H^2$ in a form
parametrized by the geodesic length $2\rho$.  If we wish to
determine the geodesic between two given time-like points $x^1 \neq x^2$,
we can determine the normal and the parametrized geodesic
from the vector  product \cite{RA} p. 64-66
$k \sim (x^1 \times x^2)$, which is perpendicular to
$x^1, x^2$, by an appropriate normalization and parametrization.

\subsection{The hyperbolic disc.}

The hyperbolic disc parametrization of $H^2$, 
compare \cite{RA} pp. 127-135, 
is given by 
the map of the hyperboloid to the interior of the unit circle,
\begin{equation}
\label{g22}
x,\;  \langle x,x \rangle=1 \rightarrow z 
= (x_1+ix_2)/(1+x_0),\; x_0 \geq 1, |z| \leq 1.
\end{equation}
with inverse
\begin{equation}
\label{g222}
x_0= (1+|z|^2)/(1-|z|^2),\; x_1+ix_2= z\;2/(1-|z|^2).
\end{equation}
The map from the hyperboloid to the hyperbolic disc is 
conformal \cite{RA} p.8,
i.e. preserves the angle between geodesics.
The hyperbolic disc represents $SU(1,1)/U(1)$  within 
$M(1,2)$. The action of Lorentz transformations on the disc is nevertheless 
described by the group $SU(1,1)$. It is given by the linear fractional
transform
\begin{eqnarray}
\label{g23}
g&=&
\left[
\begin{array}{ll}
\lambda & \mu\\
\overline{\mu}&\overline{\lambda}
\end{array} \right],
\\ \nonumber
(g,z)\rightarrow w &=& 
\frac{\lambda z + \mu}{\overline{\mu} z +\overline{\lambda}}.
\end{eqnarray}

Extend the complex variable $z$ to  the full complex plane $C$ and consider
a circle of radius $R$ centered at $q \in C$. Its equation is
\begin{equation}
\label{g24}
(\overline{z-q})(z-q)-R^2
= \overline{z}z-\overline{q}z-q\overline{z}+\overline{q}q-R^2=0.
\end{equation} 
Define  the  vector 
$k=k_0(1,  {\rm Re}(q), {\rm Im}(q)), k_0>1$.
The vector $k$ is space-like if $|q|^2 \geq 1$. For  the intersection
of the plane perpendicular to $k$ with $H^2$ we find
with eqs. \ref{g22},\ref{g222}
\begin{equation}
\label{g25}
0= \langle k,x \rangle 
= \frac{1}{2} k_0(1+x_0)(\overline{z}z-\overline{q}z-q\overline{z}+1).
\end{equation}
Comparing eq. \ref{g24} we find: The geodesic intersection of the
plane perpendicular to space-like $k$ in the hyperbolic disc model
is a circle with center $q=q(k), |q|^2 \geq 1$
and of  radius $R=\sqrt{|q|^2-1}$. Comparison with the unit circle
$|z|^2=1$ shows that the two circles have perpendicular intersections.
An example of a geodesic circle is given in Fig. 2 below. This circle 
contains an edge line of an octagon.

Since planar intersections in the hyperboloid model map into
circles in the hyperbolic disc model, we look for the action of
Weyl reflections on the hyperbolic disc. They must be reflections 
in the circles.
 
{\bf Prop 2}: The Weyl reflection with space-like unit Weyl vector 
$k, \langle k,k \rangle=-1$ in the hyperbolic disc model determines
a non-analytic fractional map
\begin{eqnarray}
\label{g26}  
q&=& (k_1+ik_2)/(k_0)=(\cosh(2\theta)/(\sinh(2\theta)) \exp(2i\alpha),
\\ \nonumber 
k&=&k_0(1, {\rm Re}(q), {\rm Im}(q)),\;
k_0=R^{-1}=1/(\sqrt{\overline{q}q-1}),
\\ \nonumber
(W_k,z) &\rightarrow& w=s_k z=g\overline{g}^{-1}\psi_1 z
=g\overline{g}^{-1}\overline{z}
\\ \nonumber
&=& 
\frac{a\overline{z}+b}{c\overline{z}+d},
\left[
\begin{array}{ll}
a & b\\
c & d\\
\end{array}
\right] =g\overline{g}^{-1},
\end{eqnarray}
where $g\overline{g}^{-1}$ is given in eq. \ref{gA3}.\\
{\em Proof}: The expression eq. \ref{g26} arises by application of the
automorphism eq. \ref{gA3}: The automorphism $\psi_1$ acting on  the hyperbolic 
disc yields 
$z \rightarrow \overline{z}$ and is followed by the fractional 
transform with the matrix given in eq. \ref{gA3}. 

In the hyperbolic disc model, any geodesic circle divides the unit
circle into two parts. These two parts are mapped into one another
under a Weyl reflection with space-like Weyl vector.
Given two points $z^1, z^2,\; z^1\neq 0$ in the hyperbolic disc model, the unique
geodesic circle which passes through both of them can be obtained by 
inserting $z^1, z^2$ into eq. \ref{g24} and solving for the center
$q$. One finds
\begin{equation}
\label{g27}
q=q(z^1,z^2)= (\overline{z}^1z^2-z^1\overline{z}^2)^{-1}
(z^2-z^1-z^1z^2\overline{(z^2-z^1)}). 
\end{equation}
In case $\overline{z}^1z^2-z^1\overline{z}^2=0$ when eq. \ref{g27}
does not apply the geodesic is a straight line 
$z = \tanh (\rho) \; z^1/|z^1|$
through the origin of the 
unit disc.

\subsection{Pseudoriemannian structure and curvature on $H^2$.}

For the notation we refer to \cite{LA} pp. 146-7.
On 2D surfaces in $M(1,2)$ with coordinates $u^{\mu}$ we define a pseudoriemannian metric with line element
\begin{eqnarray}
\label{g28}
ds^2&=& \sum_{\mu\nu} g_{\mu\nu} du^{\mu}du^{\nu}
\\ \nonumber
&=&\sum_{\mu,\nu} 
(\frac{\partial x_0}{\partial u^{\mu}}\frac{\partial x_0}{\partial u^{\nu}}
-\frac{\partial x_1}{\partial u^{\mu}}\frac{\partial x_1}{\partial u^{\nu}}
-\frac{\partial x_2}{\partial u^{\mu}}\frac{\partial x_2}{\partial u^{\nu}})
du^{\mu}du^{\nu}.
\end{eqnarray}
On $H^2$ with coordinates eq. \ref{g20} we obtain with $u^1=2\theta, u^2=2\phi$
\begin{eqnarray}
\label{g29}
g_{..} &=& (g_{\mu\nu})=
\left[
\begin{array}{ll}
-1 & 0\\
0& -(\sinh(2\theta))^2
\end{array}
\right],
\\ \nonumber
g^{..} &=& (g^{\mu\nu})=
\left[
\begin{array}{ll}
-1 & 0\\
0& -(\sinh(2\theta))^{-2}
\end{array}
\right],
\\ \nonumber
ds^2&=& -(2d\theta)^2-(\sinh(2\theta))^2d(2\phi)^2\leq 0.
\end{eqnarray}
For any curve $\theta=\rho, \phi=c_0$ on $H^2$ with parameter $\rho$
we find 
\begin{equation}
\label{g210}
(\frac{ds}{d(2\rho)})^2=-1, 
\end{equation}
so that $2\rho$ is a geodesic length parameter.
General geodesics with the same length parameter
can then be constructed as given in eq. \ref{g21}.

In general relativity, the geodesics are the world lines
followed by free test particles. Null geodesics $ds^2=0$ 
are followed by photons, and geodesics $ds^2>0$ are
world lines for massive test particles.
Note that all points on $H^2$ have space-like separation
and so can be connected only by geodesics with $ds^2<0$.

From the only non-vanishing derivative 
$\partial g_{22}/\partial u^1=-2\sinh(2\theta)\cosh(2\theta)$
we get as non-vanishing Christoffel symbols
\begin{eqnarray}
\label{g211}
\Gamma^1_{22}&=& -2\sinh(2\theta)\cosh(2\theta),\;
\Gamma^2_{12}= \Gamma^2_{21} =  \cosh(2\theta)/\sinh(2\theta).
\end{eqnarray}
The element of interest of the Riemannian curvature tensor
becomes $R_{1212}= -(\sinh(2\theta))^2$, and from it the scalar 
curvature $R$ and the Ricci tensor $R_{ij}$ become
\begin{eqnarray}
\label{g212}
R&=& -2,\; 
(R_{ij})
=
-\left[
\begin{array}{ll}
1& 0\\
0& (\sinh(2\theta))^2
\end{array}
\right].
\end{eqnarray}
So $H^2$ with the metric eq. \ref{g29}  has  constant negative curvature,
compare \cite{LA} and \cite{RA} p. 5.

\section{Topology and metric of the double torus.}

We wish to consider a 2D cosmological model with the topology 
of a double torus. The   double torus may be unfolded into an octagon 
as described by Klein \cite{KL} pp. 264-268 and 
by Hilbert and Cohn-Vossen \cite{HI} p. 265. 
This octagon 
in topology is denoted as 
the fundamental domain \cite{LA} of the double torus. The pairwise 
gluing of the octagon edges then reflects the topology. All the 
eight vertices of this octagon represent the same point of the 
double torus.
The universal covering manifold of the double torus from this 
unfolding must admit
a tesselation by octagons. Such a tesselation requires that 
eight octagons be arranged without overlap around a single vertex.
This topological condition enforces  the hyperbolic space $H^2$ 
and its geometry as the universal covering \cite{LA}
of the octagon.

\subsection{Group description of the topology.}

The topology of the double torus can be characterized by its homotopy 
group $\Phi_2$ whose elements are closed paths starting at the same point. 
This homotopy group is described in \cite{CO,HI}. It 
has four generators with a single
relation between them. Two pairs of generators are associated each to a 
single torus. The group relation, see eq. \ref{g48} below,  
arises from the gluing of the
two tori into the double torus along a glue line, compare Fig. 3 below. 

Upon embedding the octagon into its universal covering $H^2$, there must 
exist a fixpoint-free
action of the homotopy group $\Phi_2$ on $H^2$. This action was explicitly 
given by Magnus \cite{MA} and will be described in subsection 5.1. 
To any element of $\Phi_2$ there must correspond one and only one 
position of an octagon on $H^2$. The topological fundamental domain property 
of the octagon now implies that any orbit  on $H^2$ under $\Phi_2$
has one and only one  representative on a single octagon. 
This property assures the fundamental domain property from the point of
view of group action. Points on the
intersection of different octagons require an extra treatment.

We now describe the various manifolds associated to the double torus
in terms of actions, subgroups and cosets of the group SU(1,1).

Select as the fundamental domain of $\Phi_2$ 
the octagon $X_0$ centred at $x^0=(1,0,0)$ on $H^2$, corresponding to 
$z_0=0$ 
on the disc, see Fig. 2. For any point $z$ of this octagon there exists an element
$p \in SU(1,1)/U(1): z_0 \rightarrow z$. For an arbitrary point $c \sim z \in H^2$
interior to an octagon, the 
tesselation of $H^2$ by octagons as fundamental domains implies that 
there exist unique elements 
$\langle \phi \in \Phi_2,\; p \in SU(1,1)/U(1)\rangle$ 
such that 
\begin{equation}
\label{g31}
c= \phi\, p,\; \phi \in \Phi_2.
\end{equation} 

Combined with eq. \ref{g0132} we find for arbitrary elements of $SU(1,1)$
a unique factorization 
\begin{equation}
\label{g32}
g= \phi\, p\,h,\; \phi \in \Phi_2,\; h \in U(1).
\end{equation}
This is a double coset factorization of $SU(1,1)$ with respect to the left 
subgroup $\Phi_2$ and the right subgroup $U(1)$. The group elements
$p$ which generate the points of the initial octagon $X_0$ are the 
representatives of the corresponding double cosets 
$\Phi_2\backslash SU(1,1)/U(1)$.

For the group actions we obtain from eq. \ref{g32}: Under the left 
action of $\Phi_2$ we find $(\tilde{\phi},\phi\,p\,h) \rightarrow 
(\tilde{\phi}\,\phi)\, p\,h$ governed by group multiplication 
within $\Phi_2$. The general left action of $SU(1,1)$,  
$(\tilde{g},\phi\,p\,h) \rightarrow 
\phi'\, p'\,h' $ implies on $H^2$ that, starting from any fixed point of an arbitrary 
fixed octagon, we can find its unique image on a unique  image octagon.
The map $(\tilde{g}, p) \rightarrow p'$ 
depends on $\phi$ but is independent of $h'$. It 
determines a transitive  action of $SU(1,1)$ on the octagon $X_0$ 
modulo the group 
$\Phi_2$. The group $SU(1,1)$ with this action on $X_0$ is the most general one 
compatible with the pseudoriemannian metric.

\subsection{Geodesics on the double torus.}

Consider geodesics on $H^2$, equipped with the pseudoriemannien metric
of subsection 3.3. Viewing them  as sections of the hyperboloid 
in $M(1,2)$ with planes through $(0,0,0)$ perpendicular to space-like vectors \cite{RA}
one concludes that they are always infinite. Any such geodesic starting at a 
point of the initial octagon $X_0$ 
will cross a sequence of octagons. To get the geodesic on
the double torus, we must pull its points back to the initial octagon.
Both the geodesic property and the geodesic distance are unchanged under 
this pull-back.
The full geodesic on the octagon will consist of all these pull-backs.
Take the center $x^0=(1,0,0)$ of the initial octagon $X_0$ and a geodesic 
$x(\tau), x(0)=x^0$ on 
$H^2$ of fixed direction. Suppose that that geodesic on $H^2$ hits 
at $\tau=\tau'$ for the first 
time a center $x^1=x(\tau')=L(\phi)\,x(0)$ of another octagon in the tesselation. 
Then the pull-back of the geodesic to $X_0$ must hit the center 
of $X_0$ after a finite geodesic distance and therefore must 
close on $X_0$. Thus the search for closed and finite geodesics 
through the center of
the initial fundamental domain is converted   on the
hyperbolic covering manifold $H^2$ into the crystallographic search for 
sections of geodesics,
characterized by their length and direction,  
which connect the centers of octagons in the tesselation.
In what follows we shall focus on the representative closed 
geodesics passing through the center of the octagon.
In subsections 6.2 and 6.4 we shall determine 
the directions and the shortest geodesic distances between
centers of octagons. In the double torus model,
these geodesics become the shortest closed 
geodesics. As explained in the introduction and in detail in \cite{LA}, 
the shortest closed geodesics are of interest for observing the topology of the model.

We can compute the geodesic distance from $x^0$ to
the center $x^1=L(\phi)\, x^0$ of the image octagon from the group element 
$\phi \in \Phi_2$.
When the group element is written in terms of the Euler angles eq. \ref{g011}
as $\phi= g(\alpha,\theta,\gamma)$, 
the scalar product of the vectors $x_0, x_1$  pointing to the two centers is
determined by the second hyperbolic Euler angle as 
\begin{equation}
\label{g321}
\langle x^0, x^1 \rangle = 
\langle x^0,  L(\phi(\alpha,\theta,\gamma))\,x^0 \rangle= 
\cosh (2\theta),\, \theta=\theta(\phi).
\end{equation}
Any other point $y^0$ inside the octagon $X_0$ from eqs. 
\ref{g31}, \ref{g32} can be 
reached by application 
of a fixed group element $p: \; y^0=L(p)x^0$. The image $y^1$ 
of $y^0$ under $\phi \in \Phi_2$
is in the octagon whose center is $L(\phi) x^0$ and located at the point
$L(\phi p) x^0$. By construction, the geodesic from $y^0$ to its
image $y^1=\phi y^0, \phi \in \Phi_2$ is closed on the double torus.
The geodesic distance from $y^0$ to $y^1$ is
determined by the hyperbolic cosine  
\begin{equation}
\label{g33}
\langle y^0, y^1 \rangle 
=\langle x^0, L((p^{-1}\phi p)(\alpha',\theta', \gamma'))\,x^0 \rangle
=\cosh (2\theta'),\, \theta'=\theta'(p^{-1}\phi p).
\end{equation}
Here we used the invariance of the scalar product under 
the Lorentz group $SO_{\uparrow}(1,2)$.
This expression  generalizes  eq. \ref{g321}. It means that the 
search for general closed geodesics will depend
on the initial point of the octagon. In subsection 6.2 we 
shall extend the search to vertices
of the octagon. An analysis of eq. \ref{g33}
for general points requires crystallographic elaboration 
and will not be given here.

The pull-back of any geodesic on $H^2$ which does not connect 
corresponding points for any pair of octagons
does not close on $H_0$.

In the next sections we describe the details of the octagon and of the
homotopy group.

\section{The Coxeter and the homotopy group.}
We describe the homotopy group of the double torus as a subgroup
of a hyperbolic Coxeter group.

\subsection{The hyperbolic Coxeter group.}

Consider the hyperbolic Coxeter group
generated by three  Weyl reflections eq. \ref{g014}   with Weyl vectors and
relations
\begin{eqnarray}
\label{g411}
&& R_1: k=(0,0,-1),\; R_2:k=(0,\cos(3\pi/8),\sin(3\pi/8)),
\\ \nonumber 
&& R_3:k=(\sinh(2\theta), -\cosh(2\theta),0),\cosh(2\theta)=\cot(\pi/8),
\\ \nonumber
&& R_1^2=R_2^2=R_3^2=e,
\\ \nonumber 
&& (R_1R_2)^2=(R_2R_3)^8=(R_3R_1)^8=e.
\end{eqnarray}
We follow Magnus \cite{MA} pp. 81-95 who denotes the Coxeter group by $T(2,8,8)$.
Products of Weyl reflections in $M(1,2)$ generate Lorentz
transformations
\begin{equation}
\label{gA4}
(R_1R_2)=L(B),\; (R_2R_3)=L((AB)^{-1}),\; (R_3R_1)=L(A).
\end{equation} 
which Magnus \cite{MA} pp. 87-88 expresses as representations of elements $A,B \in SU(1,1)$:
\begin{eqnarray}
\label{g41}
A:&=&
i/\sin(\alpha_0)
\left[
\begin{array}{ll}
\cos(\alpha_0)& \rho\\
-\rho& -\cos(\alpha_0)
\end{array}
\right],
\\ \nonumber
B:&=&
\left[
\begin{array}{ll}
\exp(i\alpha_0)& 0\\
0& \exp(-i\alpha_0)
\end{array}
\right],
\\ \nonumber
AB:&=&
i/\sin(\alpha_0)
\left[
\begin{array}{ll}
\cos(\alpha_0)\exp(i\alpha_0)& \rho\exp(-i\alpha_0)\\
-\rho\exp(i\alpha_0)& -\cos(\alpha_0)\exp(-i\alpha_0)
\end{array}
\right],
\\ \nonumber
\alpha_0&:=&\pi/8,\; \cot(\alpha_0)=((1+\sqrt{1/2})/(1-\sqrt{1/2}))^{1/2},
\\ \nonumber
\rho &=& \sqrt{\cos(\pi/4)}= \sqrt{\sqrt{1/2}}.
\end{eqnarray}
Here the fixed angle $\alpha_0=\pi/8$ is taken from Magnus and should be 
carefully distinguished from the general Euler angle $\alpha$ used in eq. \ref{g09}.
All three matrices are of elliptic type. By use of eq. \ref{g06}
and eq. \ref{g011} we find for the exponential and Euler parameters  the 
values
\begin{eqnarray}
\label{g42}
A&:& \tilde{\alpha}=\pi/2,\, \omega(A)=(\cot(\alpha_0),-\rho/\sin(\alpha_0),0),
\\ \nonumber 
&& z(A)= \exp(i\pi)\sqrt{\cos(2\alpha_0)/(1+\sin(2\alpha_0))},
\\ \nonumber
&&\Omega(A)= (\pi/2,{\rm arccosh}(\cot(\alpha_0)),0),
\\ \nonumber
B&:& \tilde{\alpha}=\pi/8,\, \omega(B)=(1,0,0),
\,z(B)=0,
\\ \nonumber
&&\Omega(B)= (\alpha_0,0,0),
\\ \nonumber
AB&:& \tilde{\alpha}=\pi-\alpha_0,\, 
\omega(AB)=((\cot(\alpha_0))^2,-\rho\cos(\alpha_0)/(\sin(\alpha_0))^2,
\rho/\sin(\alpha_0)),
\, 
\\ \nonumber
&& z(AB)=  \exp(i7\pi/8)\sqrt{\cos(2\alpha_0)},
\\ \nonumber
&&\Omega(AB)= 
(\pi/2,{\rm arccosh}(\cot(\alpha_0)),\alpha_0).
\\ \nonumber
\end{eqnarray}

The points $z(g)=(\omega_1+i\omega_2)/(1+\omega_0), g=A, B, AB $ 
represent the fixpoints  
$\omega (g)$ in  the hyperbolic disc model.

The matrices $A,B,AB$ have the following properties which are 
relevant under the homomorphic map to the Lorentz group:
\begin{equation}
\label{g43}
A^2=-e,\; B^8=-e,\; (AB)^8=-e.
\end{equation}
The vectors $\omega(B), \omega(AB), \omega(A)$ are the vertices of the
triangular fundamental  domain on $H^2$ for the Coxeter group $T(2,8,8)$,
compare Fig. 2.
The hyperbolic edge lengths of this triangle are given from
eqs. \ref{g321} and \ref{g42}  and from hyperbolic trigonometry,
\cite{RA} p. 86, by
\begin{eqnarray}
\label{g45}
\langle \omega(B), \omega(AB) \rangle &=:& \cosh(2\theta(B,AB))
=(\cot(\alpha_0))^2,
\\ \nonumber
\langle \omega(AB), \omega(A) \rangle &=:& \cosh(2\theta(AB,A))=\cot(\alpha_0),
\\ \nonumber
\langle \omega(A), \omega(B) \rangle &=:&\cosh(2\theta(A,B))= \cot(\alpha_0),
\end{eqnarray}
The hyperbolic angle $2\theta(B,AB)$ is defined between the unit
vectors $\omega(B),\omega(AB)$.
The hyperbolic triangle has two edges of equal length which form an angle
$\pi/2$, and two more angles $\pi/8$.

\subsection{The homotopy group.}

With Magnus \cite{MA} we pass from hyperbolic triangles to octagons and
from the hyperbolic Coxeter group to the homotopy group $\Phi_2$.
The octagon, the fundamental domain of the group $\Phi_2$,
is obtained by applying all reflections in the planes containing
$\omega(B)=(1,0,0)$ and one of the vectors 
$(\omega(A), \omega(AB))$. It consists of 16 triangles. The vertices of the octagon are 
obtained from $\omega(AB)$,
the midpoints of edges from  $\omega(A)$ by applying all powers of $L(B)$
to these two vectors respectively.
The images of the octagon under
$\Phi_2$ yield a tesselation of $H^2$.

\begin{center}
\begin{picture}(0,0)%
\includegraphics{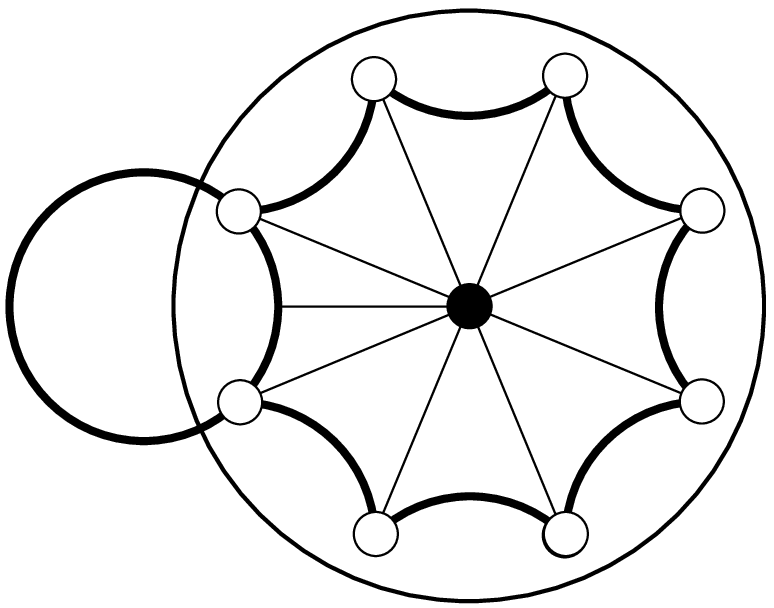}%
\end{picture}%
\setlength{\unitlength}{4972sp}%
\begingroup\makeatletter\ifx\SetFigFont\undefined%
\gdef\SetFigFont#1#2#3#4#5{%
  \reset@font\fontsize{#1}{#2pt}%
  \fontfamily{#3}\fontseries{#4}\fontshape{#5}%
  \selectfont}%
\fi\endgroup%
\begin{picture}(2913,2280)(-1772,-301)
\put(-1079,788){\makebox(0,0)[lb]{\smash{\SetFigFont{12}{14.4}{\familydefault}{\mddefault}{\updefault}
$z(A)$
}}}
\put(-629,1160){\makebox(0,0)[lb]{\smash{\SetFigFont{12}{14.4}{\familydefault}{\mddefault}{\updefault}
$z(AB)$
}}}
\put(238,784){\makebox(0,0)[lb]{\smash{\SetFigFont{12}{14.4}{\familydefault}{\mddefault}{\updefault}
$z(B)$
}}}
\put(908,1644){\makebox(0,0)[lb]{\smash{\SetFigFont{12}{14.4}{\familydefault}{\mddefault}{\updefault}
$|z|=1$
}}}
\end{picture}

\end{center}

Fig. 2. The octagon that forms the fundamental domain of the 
double torus in the unit circle $|z|=1$ of the hyperbolic disc model. 
The eight  edges are parts of geodesic
circles as indicated on the left-hand side. The vertices of the fundamental 
hyperbolic Coxeter triangle are marked by the complex numbers
$z(B), z(AB),z(A)$ which represent the rotation axes of these 
elliptic elements in the hyperbolic disc model. 
\vspace{0.2cm}

In the octagon tesselation of $H^2$, any vertex is surrounded by
eight octagons sharing that vertex. 
The Coxeter group eq. \ref{g411} 
has an involutive automorphism by the interchange 
of the generators $R_1, R_2$. As a consequence, any Coxeter triangle
has two equal angles, and the triangular tesselation has 
two classes of vertices with congruent surroundings. These classes form 
the centers and the 
vertices respectively of the octagon tesselation, compare Fig. 2.
Seen from any center of an octagon, the hyperbolic distance
of its interior points to its center is  smaller 
than their distance to any other octagon center.
So the octagons form hyperbolic Voronoi cells.
The tesselation seen not from the octagon centers but
rather from their vertex set is a copy of the octagon tesselation.
The new octagons around the vertices may be called  the dual Delone cells 
and in shape coincide with the octagons.

The homotopy group $\Phi_2$ as a subgroup of the 
Coxeter group $T(2,8,8)$ can be generated by (an even number of) 
Weyl reflections in the hyperbolic 
edges of this triangle. Center and vertices of the octagon are marked by 
full and open circles respectively. 
The homotopy group $\Phi_2$ according to Magnus \cite{MA} pp. 92-93 is a subgroup 
of the hyperbolic Coxeter group and has the generators
\begin{equation}
\label{g46}
C_0=AB^{-2},\; C_1=BC_0B^{-1},\; C_4=B^4C_0B^{-4},\; C_5=B^5C_0B^{-5}.
\end{equation}
which have the Euler angle parameters
\begin{eqnarray}
\label{g47}
C_0&:&\Omega= (\pi/2,{\rm arccosh}(\cot(\alpha_0)),-\pi/4),
\\ \nonumber
C_1&:&\Omega= 
(5\pi/8,{\rm arccosh}(\cot(\alpha_0)),-3\pi/8),
\\ \nonumber
C_4&:&\Omega= 
(\pi,{\rm arccosh}(\cot(\alpha_0)),-3\pi/4),
\\ \nonumber
C_5&:&\Omega= 
(9\pi/8,{\rm arccosh}(\cot(\alpha_0)),-7\pi/8).
\\ \nonumber
\end{eqnarray}
We call this set-up the Magnus representation of the homotopy group.
Note that the group $\Phi_2$ acts without fixpoints and so, in contrast
to the Coxeter group, does not contain reflections or rotations.

The generators are subject to the relation
\begin{equation}
\label{g48}
(C_0C_1^{-1}C_0^{-1}C_1)(C_4C_5^{-1}C_4^{-1}C_5)=e.
\end{equation}
As mentioned in subsection 4.1, this relation in topology expresses the 
gluing of the double torus from two single tori.
A geometric interpretation of this group relation
will be given in subsection 6.1.

\subsection{The Coxeter group acting on $SU(1,1)$.}

We wish to implement the Coxeter group in terms of actions
on $SU(1,1)$. Our  goal is to obtain the Magnus representation
eq. \ref{g41} 
of the homotopy group and its embedding into the Coxeter group
on the level of $SU(1,1)$ without use of the Lorentz group. 
Another reason for this analysis can be seen by comparison 
with free spin-$(1/2)$ fields in special relativity and their  discrete versions
\cite{KR3}: These spinors 
under the unimodular time-preserving Lorentz group transform according
to its covering group $Sl(2,C)$. Involutions outside this group like parity or Coxeter
reflections for them must be introduced
by automorphisms. In the present case, spinors on $H^2$ are
two-component fields $(\chi_1(x),\chi_2(x))$. A natural action 
of $g \in SU(1,1)$ on spinors is defined by 
\begin{equation}
\label{gA5a}
\left[
\begin{array}{l}
(T_g\chi_1)(x)\\
(T_g\chi_2)(x)
\end{array} \right]
:= g 
\left[
\begin{array}{l}
\chi_1(L^{-1}(g)x)\\
\chi_2(L^{-1}(g)x)
\end{array} \right].
\end{equation}
The operators $T_g$ in eq. \ref{gA5a} obey $T_{g_1}\circ T_{g_2}=T_{g_1g_2}$.

We turn now to the action of involutions on the spinors and employ
the involutive automorphisms of section 2.4. 
To get preimages of the Coxeter group acting on $SU(1,1)$ we 
use the Weyl vectors from eq. \ref{g411} and the parameters of
eq. \ref{gA1} and  
define by use of  eq. \ref{gA3} three involutive automorphisms
of $SU(1,1)$,
\begin{eqnarray}
\label{gA5}
s_1:k&=&(0,0,-1),\; 2\alpha=-\pi/2,\,\theta=0,
\\ \nonumber
s_1&=& g_1\overline{g_1}^{-1} \circ \psi_1,
\\ \nonumber
g_1(\overline{g_1})^{-1}&=& e,
\\ \nonumber
s_2:k&=&(0,\cos(3\pi/8),\sin(3\pi/8)),\; 2\alpha=3\pi/8,\,\theta=0,
\\ \nonumber
s_2&=& g_2\overline{g_2}^{-1}\circ \psi_1,
\\ \nonumber
g_2(\overline{g_2})^{-1}&=& r_3(7\pi/8),
\\ \nonumber
s_3:k&=&(\sinh(2\theta),-\cosh(2\theta), 0),\; 2\alpha=\pi,\,
\cosh(2\theta)=\cot(\pi/8),
\\ \nonumber
s_3&=&g_3\overline{g_3}^{-1} \circ \psi_1,
\\ \nonumber
g_3(\overline{g_3})^{-1}
&=& r_3(\pi/2)b_2(2\theta)r_3(\pi/2)r_3(\pi/2)=-r_3(\pi/2)b_2(2\theta)
\\ \nonumber
&=& 
\left[
\begin{array}{ll}
-i\cosh(2\theta)& -i\sinh(2\theta)\\
i\sinh(2\theta)& i\cosh(2\theta)
\end{array}
\right]
\end{eqnarray}
We compute  the pairwise products of these automorphisms.
The automorphism $\psi_1$ in the products cancel in pairs due to eq. \ref{gA22}
and we obtain
operators acting only by left multiplication.
We find by comparison with the 
matrices eq. \ref{g41} in the Magnus representation the left actions

\begin{eqnarray}
\label{gA6}
(s_1\circ s_2)&=& \psi_1 \circ r_3(7\pi/8)\circ \psi_1
=-r_3(\pi/8)\circ \psi_1^2=-r_3(\pi/8)
\\ \nonumber 
&=& -B,\; (s_2\circ s_3)= (AB)^{-1},\; (s_3 \circ s_1)=-A.
\end{eqnarray}

The operator relations of eq. \ref{gA6} 
lift into the relations eq. \ref{gA4}  of the Coxeter group 
due to 
$L(g)=L(-g)$. 

{\bf Prop 3}: The involutive automorphisms $s_1,s_2,s_3$ of
eq. \ref{gA5} with $L(s_i):=R_i,\, i=1,2,3$ generate
for the Coxeter group eq. \ref{g411} a preimage which acts  on $SU(1,1)$.
We shall speak about the Coxeter group generated by 
$\langle s_1,s_2,s_3 \rangle$.

\subsection{Conjugation of  $\Phi_2$ by the Coxeter group.}

The matrices $A,B$ under conjugation with the three automorphisms 
$s_1,s_2,s_3$ of eq. \ref{gA5} transform as
\begin{equation}
\label{gA7}
\begin{array}{llll}
X&s_1Xs_1&s_2Xs_2&s_3Xs_3\\
A&-A&-B^{-1}AB&-A\\
B&B^{-1}&B^{-1}&-AB^{-1}A\\
\end{array}
\end{equation}

We shall show that the homotopy group $\Phi_2$, augmented by the element
$B^4$, is a normal subgroup of the Coxeter group. The element 
$B^4, B^{-4}=-B^4$ has a simple geometric interpretation: In $M(1,2)$ it is a rotation by 
$2\tilde{\alpha}=\pi$ and transforms the two 
tori of the double torus into one another.  
Writing the generators eq. \ref{g46} of $\Phi_2$ in terms of the matrices $A,B$ one
finds the conjugation relations
\begin{equation}
\label{gA8}
B^4C_0B^{-4}=C_4,B^4C_1B^{-4}=C_5,B^4C_4B^{-4}=C_0,B^4C_5B^{-4}=C_1
\end{equation}
The subgroup generated by $B^4$ consists of the elements 
$\langle e,-e,B^4,-B^4\rangle$. Only the identity element is shared with
$\Phi_2$.
We denote the  group generated by $\langle C_0,C_1,C_4,C_5,B^4\rangle $ as
$\tilde{\Phi}_2$. It is a semidirect product with $\Phi_2$ 
as the normal subgroup.
Now we study the conjugation of the generators of this group under
the three automorphisms eq. \ref{gA5} and obtain by use of 
eqs. \ref{g46}, \ref{gA7}, \ref{gA8}:
\begin{equation}
\label{gA9}
\begin{array}{llll}
i&s_1C_is_1&s_2C_is_2&s_3C_is_3\\
0&-C_0B^4  &C_5^{-1} &-C_0^{-1}\\
1&C_5^{-1} &C_4^{-1} &-C_0C_1C_4^{-1}B^4\\
4&-C_4B^4  &C_1^{-1} &C_0C_4^{-1}C_0^{-1}\\
5&C_1^{-1} &C_0^{-1} &C_0C_5C_4^{-1}B^4\\
\end{array}
\end{equation}
We also find from eq. \ref{gA7}
\begin{equation}
\label{gA10}
s_1B^4s_1=-B^4,\;s_2B^4s_2=-B^4,\;s_3B^4s_3=-C_0C_4^{-1}B^4.
\end{equation}
All these conjugations yield again elements of the group $\tilde{\Phi}_2$.
The inclusion of $B^4$ is necessary since it appears in the 
conjugations of eq. \ref{gA9}. 
Therefore we find

{\bf Prop 4}: The extension $\tilde{\Phi}_2$ of the homotopy group
$\Phi_2$ is a normal subgroup of the Coxeter group generated by
$\langle s_1,s_2,s_3 \rangle$.

Corresponding relations hold true on the level of the Lorentz group.
In view of this property under conjugation we can call the hyperbolic 
Coxeter group the discrete symmetry group of the octagonal tesselation
associated with the double torus. 

\section{The homotopy group and closed geodesics.}
With the information on the homotopy group, its action and symmetry 
obtained in
section  5 we return to the octagonal 
tesselation and analyse the closed geodesics
as outlined in section 4.

\subsection{The action of the generators of $\Phi_2$.}

By application of the four generators of $\Phi_2$ and their inverses
to an initial octagon $X_0$ whose center is located at $z=0$,
we obtain eight images. It turns out that all of them are different
edge  neighbours of $X_0$. 
For $C_0=AB^{-2}$ we have $\theta(C_0)=\theta(A)$ and from
eq. \ref{g42}
$\cosh(\theta(A))= \cot(\alpha_0)$. From these expressions
we obtain for the geodesic distance eq. \ref{g321}
\begin{equation}
\label{g49}
\langle x^0, L(C_0) x^0 \rangle
=\cosh(2\theta(C_0))= 2(\cot(\alpha_0))^2-1
\end{equation} 
Since $L(A)$ is a rotation by $\pi$ in the midpoint of
an edge of the central octagon, the image under $C_0$ is
an edge-neighbour. The other generators eq. \ref{g46}   of $\Phi_2$ are 
conjugates of $C_0$ by powers of $B$. Therefore all  the corresponding 
images of the central octagon under the generators are edge-neighbours.
All pairs $X_0, L(\phi) X_0$ of octagons are shown in 
Figure 3. In Figs. 3 and 4 we omit the hyperbolic geometry of the
octagons as displayed in Fig. 2.

\begin{center}
\begin{picture}(0,0)%
\includegraphics{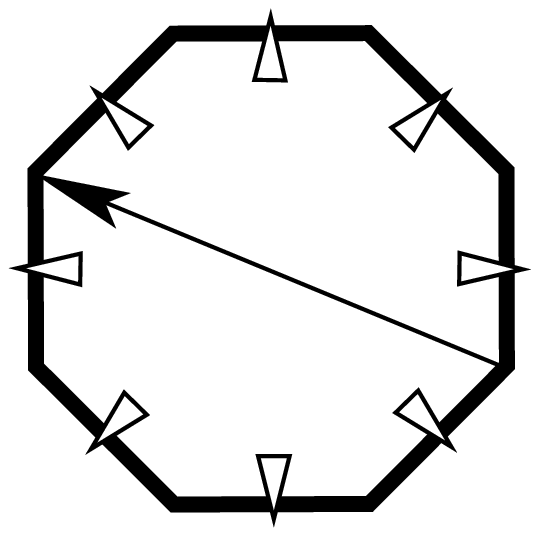}%
\end{picture}%
\setlength{\unitlength}{9945sp}%
\begingroup\makeatletter\ifx\SetFigFont\undefined%
\gdef\SetFigFont#1#2#3#4#5{%
  \reset@font\fontsize{#1}{#2pt}%
  \fontfamily{#3}\fontseries{#4}\fontshape{#5}%
  \selectfont}%
\fi\endgroup%
\begin{picture}(1500,1628)(-892,54)
\put(477,1224){\makebox(0,0)[lb]{\smash{\SetFigFont{12}{14.4}{\familydefault}{\mddefault}{\updefault}
$C_5$
}}}
\put(608,778){\makebox(0,0)[lb]{\smash{\SetFigFont{12}{14.4}{\familydefault}{\mddefault}{\updefault}
$C_4$
}}}
\put(428,324){\makebox(0,0)[lb]{\smash{\SetFigFont{12}{14.4}{\familydefault}{\mddefault}{\updefault}
$C_1^{-1}$
}}}
\put(-173,1487){\makebox(0,0)[lb]{\smash{\SetFigFont{12}{14.4}{\familydefault}{\mddefault}{\updefault}
$C_4^{-1}$
}}}
\put(-892,774){\makebox(0,0)[lb]{\smash{\SetFigFont{12}{14.4}{\familydefault}{\mddefault}{\updefault}
$C_0$
}}}
\put(-682,324){\makebox(0,0)[lb]{\smash{\SetFigFont{12}{14.4}{\familydefault}{\mddefault}{\updefault}
$C_1$
}}}
\put(-764,1225){\makebox(0,0)[lb]{\smash{\SetFigFont{12}{14.4}{\familydefault}{\mddefault}{\updefault}
$C_5^{-1}$
}}}
\put(-177, 54){\makebox(0,0)[lb]{\smash{\SetFigFont{12}{14.4}{\familydefault}{\mddefault}{\updefault}
$C_0^{-1}$
}}}
\end{picture}

\end{center}

Fig. 3. Schematic view of the octagon, the fundamental domain of the
double torus. The images of the octagon under the  four generators 
$C_i$ of the homotopy group $\Phi_2$ and 
their inverses $C_i^{-1}$ are eight edge neighbours. Outward pointing arrows 
mark the action of these generators.
The double torus can be glued from two single tori whose homotopy 
groups have the
generators $C_0,\,C_1$ and  $C_4,\,C_5$ respectively. The gluing
of these two tori inside the octagon is marked by an arrowed line.
\vspace{0.2cm}

The action of the generators of $\Phi_2$  described so far 
refers to the full hyperbolic
disc as the covering manifold of the double torus. 
The relation to the homotopy group as described by 
closed paths on the double torus 
arises as follows: Take a
homotopic path $P$ which passes from $X_0$ into $L(\phi) X_0$ through a shared 
edge. Mark first this shared entrance edge as seen from the center of 
$L(\phi) X_0$.
Then identify the preimage of this entrance edge seen from the center
of $X_0$. The reentry path P into $X_0$ passes through this 
edge. This information is presented in Figure 3. The path in
the homotopy group for any generator may in fact be represented 
on the covering manifold by a
geodesic section between two octagons that share an edge, compare Fig. 4.

We would like to use the information on the action of
single generators to find  products 
of generators which, when applied to the central octagon,
give a sequence of  pairwise edge-sharing images of $X_0$ around a vertex. 

The action of a generator
$g_i=C_j$ is always obtained from products of Weyl reflections in the edges
of the fixed Coxeter triangle in the central
octagon shown in Fig. 2. We call it a passive transformation. 
Consider a word $g= g_1g_2g_3...$ and rewrite it by use of 
conjugations as
\begin{eqnarray}
\label{g410}
g= g_1g_2g_3 \ldots &=&\ldots((g_1g_2)g_3(g_1g_2)^{-1})(g_1g_2g_1^{-1})g_1
\\ \nonumber
&=& \ldots g_3^*g_2^*g_1^*.
\end{eqnarray}  
The word on the left-hand side is  expressed on the right-hand side 
as a product of conjugates $g_j^*$ whose terms
appear in reverse order.  When acting on the initial central octagon
say $X_0$, 
the term $g_i^*$ conjugate to $g_i$ on the right-hand side 
acts on the image $L(g_1g_2...g_{i-1})X_0$ of $X_0$. Therefore the right-hand side
of eq. \ref{g410} expresses a sequence of active transformations,
and we learn that these active transformations 
appear in the reverse order of  the sequence of
passive transformations.

We apply the active transformations to the passage around the selected
vertex $z= {\rm coth}(\theta_0) \exp(i7\pi/8)$ of $X_0$.
We use the geometric information shown in Figs. 3,4
to pass  counterclockwise around this vertex through a 
sequence of edges. 
We find the sequences
\begin{eqnarray}
\label{411}
&&C_5^{-1}, C_4C_5^{-1}, C_5 C_4C_5^{-1}, 
C_4^{-1}C_5 C_4C_5^{-1}, C_1^{-1}C_4^{-1}C_5 C_4C_5^{-1},
\\ \nonumber
&&C_0C_1^{-1}C_4^{-1}C_5 C_4C_5^{-1},
C_1C_0C_1^{-1}C_4^{-1}C_5 C_4C_5^{-1},
C_0^{-1}C_1C_0C_1^{-1}C_4^{-1}C_5 C_4C_5^{-1}.
\end{eqnarray}
which should finally restore the octagon $X_0$.
The full passive sequence from eq. \ref{411} is 
\begin{equation}
\label{412}
g=C_5^{-1}C_4C_5C_4^{-1}C_1^{-1}C_0C_1C_0^{-1}=e
\end{equation}

It becomes  the identity by using the single relation eq. \ref{g48}
of the group $\Phi_2$. We have found a geometric interpretation 
of this relation on $H^2$: the subwords of this
relation in the Magnus representation describe active transformations of 
the octagon around one of its vertices.
Similar results apply to all eight vertices, and we list them 
in Table 1. Moreover we compute the geodesic distance from the 
central octagon to these neighbours in Table 2.

\begin{center}
\begin{picture}(0,0)%
\includegraphics{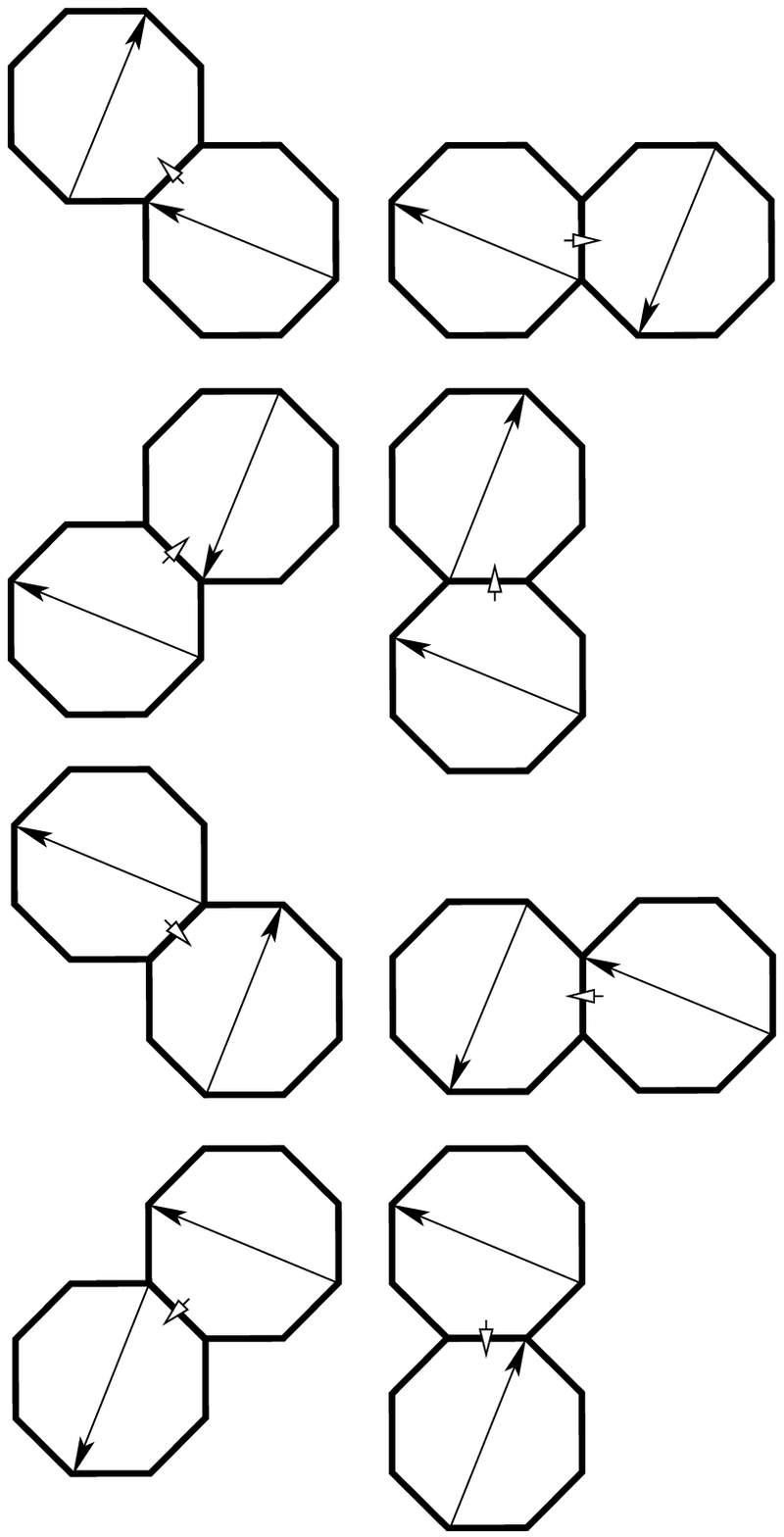}%
\end{picture}%
\setlength{\unitlength}{4144sp}%
\begingroup\makeatletter\ifx\SetFigFont\undefined%
\gdef\SetFigFont#1#2#3#4#5{%
  \reset@font\fontsize{#1}{#2pt}%
  \fontfamily{#3}\fontseries{#4}\fontshape{#5}%
  \selectfont}%
\fi\endgroup%
\begin{picture}(5078,7291)(-2275,-5327)
\put(-2245,747){\makebox(0,0)[lb]{\smash{\SetFigFont{12}{14.4}{\familydefault}{\mddefault}{\updefault}
$C_5^{-1}$
}}}
\put(2803,747){\makebox(0,0)[lb]{\smash{\SetFigFont{12}{14.4}{\familydefault}{\mddefault}{\updefault}
$C_4$
}}}
\put(2803,-843){\makebox(0,0)[lb]{\smash{\SetFigFont{12}{14.4}{\familydefault}{\mddefault}{\updefault}
$C_4^{-1}$
}}}
\put(-2245,-850){\makebox(0,0)[lb]{\smash{\SetFigFont{12}{14.4}{\familydefault}{\mddefault}{\updefault}
$C_5$
}}}
\put(2803,-2658){\makebox(0,0)[lb]{\smash{\SetFigFont{12}{14.4}{\familydefault}{\mddefault}{\updefault}
$C_0$
}}}
\put(2803,-4435){\makebox(0,0)[lb]{\smash{\SetFigFont{12}{14.4}{\familydefault}{\mddefault}{\updefault}
$C_0^{-1}$
}}}
\put(-2275,-2755){\makebox(0,0)[lb]{\smash{\SetFigFont{12}{14.4}{\familydefault}{\mddefault}{\updefault}
$C_1^{-1}$
}}}
\put(-2245,-4524){\makebox(0,0)[lb]{\smash{\SetFigFont{12}{14.4}{\familydefault}{\mddefault}{\updefault}
$C_1$
}}}
\end{picture}

\end{center}

Fig. 4. Schematic view  of the action of the generators of the 
homotopy group $\Phi_2$ on the central octagon. The arrowed glue line, compare 
Fig. 3, is marked inside each octagon. The image of the central octagon
under any  generator or its inverse is an edge neighbour. The direction 
of the map is indicated by a white arrow passing through the edge.
A homotopic path on the covering manifold could be taken as a geodesic
connecting the centers of octagons sharing an edge.
The sequence of generators as shown, multiplied from left to right,
produces a sequence of images which share a vertex. 
\vspace{0.2cm}

{\bf Table} 1: The words $w_{\nu}$ which transform the octagon into images sharing a vertex
$z_{\nu}= \sqrt{\cos(2\alpha)} \exp(i(2\nu-1)\pi/8)$.
The full words yield the identity element due to the group
relation eq. \ref{g48}. The seven images of the octagon around the chosen vertex 
appear in counterclockwise order if each  word is cut after 
$1,2, \ldots, 7$ entries, counted from left to right. 

\[
\begin{array}{ll}
\hline \nu & w_{\nu}\\ \hline 
0   & C_1^{-1}C_0C_1C_0^{-1}C_5^{-1}C_4C_5C_4^{-1}\\
1   & C_4C_5C_4^{-1} C_1^{-1}C_0C_1C_0^{-1}C_5^{-1}\\
2   & C_5C_4^{-1} C_1^{-1}C_0C_1C_0^{-1}C_5^{-1}C_4\\
3   & C_4^{-1} C_1^{-1}C_0C_1C_0^{-1}C_5^{-1}C_4C_5\\
4   & C_5^{-1}C_4C_5C_4^{-1} C_1^{-1}C_0C_1C_0^{-1}\\
5   & C_0C_1C_0^{-1}C_5^{-1}C_4C_5C_4^{-1} C_1^{-1}\\
6   & C_1C_0^{-1}C_5^{-1}C_4C_5C_4^{-1} C_1^{-1}C_0\\
7   & C_0^{-1}C_5^{-1}C_4C_5C_4^{-1} C_1^{-1}C_0C_1\\ \hline
\end{array}
\]

\subsection{Vertex neighbours.}

We shall use the self-dual  property of the octagon tesselation 
explained in subsection 5.2 as follows:
Instead of analyzing eight octagon centers around a single vertex we can
equally well consider eight vertices around a single center. 
Instead of 
geodesics passing through octagon centers 
we now choose those passing through octagon vertices.

Choosing a geodesics starting at a fixed vertex,
we may require it to run inside the octagon and hit another vertex.
In this way we can avoid any pull-back for these shortest 
geodesics. These geodesics are dual to the geodesics between the centers
of octagons sharing a vertex.

The positions of the eight vertices seen from the center $z=0$,
see Fig. 2, 
are given from eqs. \ref{g41},\ref{g42} by
\begin{eqnarray}
\label{g413}
z_{\nu}& =&  {\rm tanh}(\theta(AB)) \exp(i\pi(2\nu-1)/8),
\\ \nonumber 
&=& \sqrt{\cos(2\alpha)} \exp(i\pi(2\nu-1)/8),
\; \nu=1,\ldots, 7.  
\end{eqnarray}
From eq. \ref{g27} we obtain by the insertions 
$z_1\rightarrow z_0, z_2\rightarrow z_{\nu}$ for the 
center of the geodesic circle that connects $z_0$ with
$z_{\nu},\, \nu \neq 0$:
\begin{equation}
\label{g4131}
q(z_0,z_{\nu})= \exp(i\pi(\nu-1)/8)\; {\rm cotanh}(2\theta(AB))/\cos(\pi\nu/8)
\end{equation}

We compute the hyperbolic cosine for all pairs of vertices
and find from hyperbolic trigonometry, compare \cite{RA} pp. 83-91,
\begin{eqnarray}
\label{g414}
\cosh(2\theta(P_0,P_{\nu}))
&=& \cosh (2\theta(P_0))\cosh (2\theta(P_{\nu}))
\\ \nonumber
&&-\sinh (2\theta(P_0))\sinh (2\theta(P_{\nu}))\cos(\nu\pi/4)
\\ \nonumber
&=& (\cot(\alpha))^4(1-\cos(\nu\pi/4))+\cos(\nu\pi/4)
\end{eqnarray}
\begin{center}
\begin{picture}(0,0)%
\includegraphics{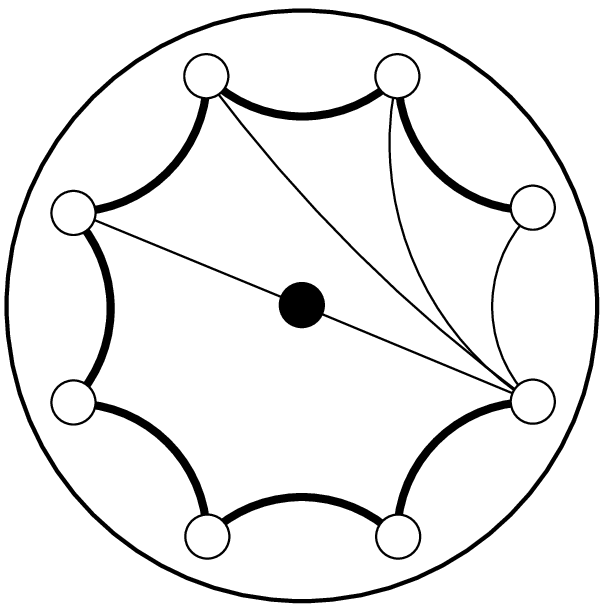}%
\end{picture}%
\setlength{\unitlength}{4972sp}%
\begingroup\makeatletter\ifx\SetFigFont\undefined%
\gdef\SetFigFont#1#2#3#4#5{%
  \reset@font\fontsize{#1}{#2pt}%
  \fontfamily{#3}\fontseries{#4}\fontshape{#5}%
  \selectfont}%
\fi\endgroup%
\begin{picture}(2280,2280)(-1259,-4005)
\put(720,-2043){\makebox(0,0)[lb]{\smash{\SetFigFont{12}{14.4}{\familydefault}{\mddefault}{\updefault}
$|z|=1$
}}}
\end{picture}

\end{center}

Fig. 5. Examples of  four shortest closed geodesics in the octagon model
as part of geodesic 
circles between vertices
(open circles) of the central octagon. They start  at the vertex 
$\nu=0$ and  end at the vertices $\nu=1,2,3,4$ respectively.
For $\nu=1$ the geodesic is an edge of the octagon. It corresponds
dually to a geodesic between centers of edge neighbours.
\vspace{0.2cm}

{\bf Table 2}: Hyperbolic cosine for geodesic distances of 
vertices of a single octagon.
\vspace{0.2cm} 

\[\begin{array}{lllll}
\hline \nu  &\nu\pi/4&\cos(\nu\pi/4)& \cosh(2\theta(P_0,P_{\nu}))
&{\rm multiplicity} \\ \hline
0    &0&1&1&         \\
1,7  &\pm\pi/4&\sqrt{1/2}&(1-\sqrt{1/2})^{-2}(-1/2+2\sqrt{1/2})&8\\
2,6  &\pm\pi/2&0&(1-\sqrt{1/2})^{-2}/(3/2+2\sqrt{1/2})&16\\
3,5  &\pm3\pi/4&-\sqrt{1/2}&  (1-\sqrt{1/2})^{-2}(7/2+2\sqrt{1/2})&16 \\
4    &\pm\pi&-1& (1-\sqrt{1/2})^{-2}(3/2+6\sqrt{1/2})&8 \\ \hline

\end{array}
\]
\vspace{0.2cm} 

The multiplicity counts geodesic sections  of equal lenghth
and starting point but of different directions which result from the 
octagonal symmetry.  
The cases $\nu=1,7$ yield the same distance as  between 
centers of edge neighbours. 
The point $P_4$
results from a rotation by $\pi$ and yields the largest 
hyperbolic distance corresponding to  a vertex neighbour.

We claim that the four octagons closest to the central octagon are
four vertex neighbours. Therefore they determine the shortest 
closed geodesics of the double torus.

\subsection{Products of generators.}

We consider products of the generators $C_i$ eq. \ref{g46}.
All possible products of two of them up to some inversions can be written in
terms of the matrices $A,B$ eq. \ref{g41} according to the following {\bf Table 3}.
\vspace{0.2cm} 

{\bf Table 3}: Products of pairs of generators of $\Phi_2$ up to inversion.
 
\[\begin{array}{llll}
\hline i,j & C_iC_j&C_i(C_j)^{-1}& (C_i)^{-1}C_j\\ \hline

0,1 & AB^{-1}AB^{-3}&-ABAB^{-1}&-B^2ABAB^{-3}\\

1,0 & BAB^{-3}AB^{-2}&-BAB^{-1}A &-B^3AB^{-1}AB^{-2}\\

0,4 &AB^2AB^{-6}&-AB^4AB^{-4}& -B^2AB^4AB^{-6}\\

4,0 &B^4AB^{-6}AB^{-2}&-B^4AB^{-4}A &-B^6AB^{-4}AB^{-2}\\

0,5 &AB^3AB^{-7}&-AB^5AB^{-5}&-B^2AB^5AB^{-7}\\

5,0 &B^5AB^{-7}AB^{-2}&-B^5AB^{-5}A &-B^7AB^{-5}AB^{-2}\\

1,4 &BABAB^{-6}&-BAB^3AB^{-4}&-B^3AB^3AB^{-6}\\

4,1 &B^4AB^{-5}AB^{-3}&-B^4AB^{-3}AB^{-1} &-B^6AB^{-3}AB^{-3}\\

1,5 &BAB^2AB^{-7}&-BAB^4AB^{-5}&-B^3AB^4AB^{-7}\\

5,1 &B^5AB^{-6}AB^{-3}&-B^5AB^{-4}AB^{-1} &-B^7AB^{-4}AB^{-3}\\

4,5 &B^4AB^{-1}AB^{-7}&-B^4ABAB^{-5}&-B^6ABAB^{-7}\\

5,4 &B^5AB^{-3}AB^{-6}&-B^5AB^{-1}AB^{-4} &-B^7AB^{-1}AB^{-6}\\

0,0 &AB^{-2}AB^{-2}&e&e\\
1,1 &BAB^{-2}AB^{-3}&e&e\\
4,4 &B^4AB^{-2}AB^{-6}&e&e\\
5,5 &B^5AB^{-2}AB^{-7}&e&e \\ \hline
\end{array}
\]
\vspace{0.2cm}

\subsection{Geodesic distances for products of generators.}

We wish to have a geometric interpretation for the action 
of products of generators on the octagon.
Consider the product $C_iC_j$ of two generators. If we define 
$C_jX_0:=X_j$ as a reference octagon we have $X_0=C_j^{-1}X_j,\;
(C_iC_j)X_0=C_iX_j$. Therefore we find that $X_0$ and $(C_iC_j)X_0$ are edge 
neighbours
to the single octagon $X_j$. When running through all products of
generators we run through all pairs of different edge neighbours to a single 
octagon. We expect to find only four different geodesic distances 
between such pairs.

We now compute for the products of generators the geodesic distances
from the matrix products of {\bf Table 3}. It suffices to compute
the second hyperbolic Euler angle for the matrices 
$AB^{\mu}A,\; \mu=1, \ldots, 7$
since all the group elements in {\bf Table 2} arise from $AB^{\mu}A$ 
by right- and left-multiplication with powers of $B$.
To these triple products we apply eqs. \ref{g012},\ref{g013} and use 
the Euler angle parameters from eq. \ref{g42}. First we
find from $AB^{\mu}A$ for the angle corresponding to  $2\gamma_1$  
in eq. \ref{g012}: $\cos(2\gamma_1)=\cos(\mu\pi/4+\pi)$
with values given in  column 3 in {\bf Table 4}.
From eq. \ref{g013} and eq. \ref{g42} we then get 
\begin{eqnarray}
\label{g415}
\cosh(2\theta(AB^{\mu}A))
&=& (\cosh(2\theta(A)))^2+(\sinh(2\theta(A))^2\cos(\mu\pi/4+\pi)
\\ \nonumber 
&=& (2(\cot(\alpha))^2-1)^2(1+\cos(\mu\pi/4+\pi))-\cos(\mu\pi/4+\pi)
\end{eqnarray}
which gives {\bf Table 4}.

{\bf Table 4}: Hyperbolic cosine for geodesic distances of 
octagon images under the products of {\bf Table 3}.
\vspace{0.2cm} 

\[\begin{array}{llll}
\hline  \mu  &\mu\pi/4+\pi&\cos(\mu\pi/4+\pi)&\cosh(2\theta(AB^{\mu}A))  \\ \hline
0    &\pi&-1&1         \\
1,7  &5\pi/4,3\pi/4&-\sqrt{1/2}&(1-\sqrt{1/2})^{-2}(3/2+2\sqrt{1/2})\\
2,6  &3\pi/2,5\pi/2&0&(1-\sqrt{1/2})^{-2} (11/2+6\sqrt{1/2})\\
3,5  &7\pi/4,9\pi/4&\sqrt{1/2}&   (1-\sqrt{1/2})^{-2}(19/2+10\sqrt{1/2})\\
4    &2\pi&1&\cosh(4\theta(A))
=(1-\sqrt{1/2})^{-2}(19/2+14\sqrt{1/2}) \\ \hline
\end{array}
\]
\vspace{0.2cm} 

From the expressions in {\bf Table 1} we can infer in particular 
all the products of
two generators which move an octagon counterclockwise around a vertex. 
They are: \\$C_1^{-1}C_0,C_0C_1,C_1C_0^{-1}, C_0^{-1}C_5^{-1},
C_5^{-1}C_4, C_4C_5, C_5C_4^{-1}, C_4^{-1}C_1^{-1}$. All of them
from {\bf Table 3} correspond to $\mu=7$.
We therefore expect that these cases in {\bf Table 4}
yield the same geodesic distance as  $\nu=2$ in {\bf Table 2} which is the case.  
The case $\mu=4$ in {\bf Table 4} gives twice  the geodesic distance of an edge neighbour,
the centers of the three octagons are on a single geodesic. 
The pull-back of this geodesic section runs twice over the shortest closed geodesic.

\section{Conclusion and Outlook.}
The double torus provides a model for a 2D octagonal cosmos with a topology of
genus 2. We study the geometry of this model by passing to its 
universal covering, the hyperbolic space $H^2$. Insight into the model
is gained from the action of groups and subgroups.
The continuous group $SU(1,1)$ acts on $H^2$ and, modulo 
the homotopy group $\Phi_2$,  on the octagon.
This action determines  the  homogeneity of both spaces. 
A discrete hyperbolic
Coxeter group acts as a subgroup on $SU(1,1)$. 
An extension of the  homotopy
group $\Phi_2$ of the double torus is a normal subgroup of this
hyperbolic Coxeter group. So the hyperbolic Coxeter group 
expresses  by conjugation the discrete  symmetry of the 
octagonal tesselation of $H^2$.
Representative  closed geodesics of the double torus model have on $H^2$ 
the crystallographic 
interpretation of geodesic links between octagon centers or vertices. 
Four shortest geodesic sections  connect pairs of vertices of the octagon.
The directions and geodesic distances
for these and the next closed geodesics are explicitly computed.

\section{Acknowledgments.}
Helpful discussions with  B. Mühlherr from the University Dortmund, Germany 
on hyperbolic Coxeter groups are gratefully acknowledged.
One of the authors (M.L.) wants to express his gratitude to the 
Director of the Institute for Theoretical Physics at the University 
of Tübingen, and to the Ministerio de Educacion in Spain for 
financial support (Grant BFM 2000-0357).

\newpage

\end{document}